\documentclass[12pt]{article}
\usepackage{amsmath,amssymb,amsfonts,amsbsy}
\usepackage{cite}
\begin{document}
\addcontentsline{toc}{subsection}{{Conservation laws and covariant equations of motion for spinning particles}\\
{\it Yu.N. Obukhov}}

\setcounter{section}{0}
\setcounter{subsection}{0}
\setcounter{equation}{0}
\setcounter{figure}{0}
\setcounter{footnote}{0}
\setcounter{table}{0}

\begin{center}
\textbf{CONSERVATION LAWS AND COVARIANT EQUATIONS OF MOTION FOR SPINNING PARTICLES}

\vspace{5mm}

\underline{Yu.N. Obukhov}$^{\,1\,\dag}$ and D. Puetzfeld$^{\,2\,\ddag}$

\vspace{5mm}

\begin{small}
(1) \emph{IBRAE, Russian Academy of Sciences, B.Tulskaya 52, 115191 Moscow, Russia} \\
  (2) \emph{ZARM, University of Bremen, Am Fallturm, 28359 Bremen, Germany} \\
  $\dag$ \emph{E-mail: obukhov@ibrae.ac.ru}\ 
  $\ddag$ \emph{E-mail: dirk.puetzfeld@zarm.uni-bremen.de}
\end{small}
\end{center}

\vspace{0.0mm} 

\begin{abstract}
We derive the Noether identities and the conservation laws for general gravitational models with arbitrarily interacting matter and gravitational fields. These conservation laws are used for the construction of the covariant equations of motion for test bodies with minimal and nonminimal coupling.
\end{abstract}

\vspace{7.2mm}

{\bf Metric-affine gravity} \cite{Obukhov:Hehl:1995} provides a general framework for the discussion of dynamics of arbitrarily interacting matter and gravitational field. In this formalism, one can analyse minimal coupling of matter with or without microstructure, along with extended nonminimal coupling schemes, in any spacetime geometry. The gravitational field potentials are the independent metric tensor $g_{ij}$ and the linear connection $\Gamma_{ki}{}^j$. The corresponding field strengths \cite{Obukhov:Hehl:1995} are the curvature, the torsion, and the nonmetricity:
\begin{eqnarray}
R_{kli}{}^j &=& \partial_k\Gamma_{li}{}^j - \partial_l\Gamma_{ki}{}^j + \Gamma_{kn}{}^j \Gamma_{li}{}^n - \Gamma_{ln}{}^j\Gamma_{ki}{}^n,\label{curv}\\
T_{kl}{}^i &=& \Gamma_{kl}{}^i - \Gamma_{lk}{}^i,\label{tors}\\ \label{nonmet}
Q_{kij} &=& -\,\nabla_kg_{ij} = - \partial_kg_{ij} + \Gamma_{ki}{}^lg_{lj} + \Gamma_{kj}{}^lg_{il}.
\end{eqnarray}
The deviation from Riemannian geometry (specified by the Christoffel connection $\widetilde{\Gamma}_{kj}{}^i = {\frac 12}g^{il}(\partial_jg_{kl} + \partial_kg_{lj} - \partial_lg_{kj})$ and marked by tilde) is measured by the {\it distorsion} tensor
\begin{equation}
N_{kj}{}^i = \widetilde{\Gamma}_{kj}{}^i - \Gamma_{kj}{}^i.\label{dist}
\end{equation}

{\bf Noether identities} arise from the symmetries of the action $I = \int d^4x {\cal L}$. Here we study the case when the Lagrangian density ${\cal L} = {\cal L}(\psi^A, \nabla_i\psi^A, g_{ij}, R_{kli}{}^j, T_{kl}{}^i, Q_{kij}, N_{kj}{}^i)$ is a function of the metric, the curvature (\ref{curv}), the torsion (\ref{tors}), the nonmetricity (\ref{nonmet}), the matter field $\psi^A$, and its {\it covariant derivative} $\nabla_k\psi^A = \partial_k\psi^A -\Gamma_{ki}{}^j\,(\sigma^A{}_B)_j{}^i\,\psi^B$. We assume that the action is invariant under general coordinate transformations of the gravitational and the matter fields: $x^i\rightarrow x^i + \delta x^i$, $g_{ij}\rightarrow g_{ij} + \delta g_{ij}$, $\Gamma_{ki}{}^j\rightarrow \Gamma_{ki}{}^j + \delta\Gamma_{ki}{}^j$, and $\psi^A \rightarrow \psi^A + \delta\psi^A$ 
\begin{eqnarray}
\delta x^i &=& \xi^i(x),\label{dex}\\ \label{dgij}
\delta g_{ij} &=& -\,(\partial_i\xi^k)\,g_{kj} - (\partial_j\xi^k)\,g_{ik},\\
\delta\psi^A&=&-\,(\partial_i\xi^j)\,(\sigma^A{}_B)_j{}^i\,\psi^B, \label{dpsiA}\\
\delta\Gamma_{ki}{}^j &=&  -\,(\partial_k\xi^l)\,\Gamma_{li}{}^j - (\partial_i\xi^l)\,\Gamma_{kl}{}^j + \,(\partial_l\xi^j)\,\Gamma_{ki}{}^l - \partial^2_{ki}\xi^j.\label{dG}
\end{eqnarray}
The generators $(\sigma^A{}_B)_j{}^i$ of the coordinate transformations satisfy commutation relations
\begin{eqnarray}
(\sigma^A{}_C)_j{}^i(\sigma^C{}_B)_l{}^k - (\sigma^A{}_C)_l{}^k (\sigma^C{}_B)_j{}^i
= (\sigma^A{}_B)_l{}^i\,\delta^k_j - (\sigma^A{}_B)_j{}^k \,\delta^i_l.\label{comms}
\end{eqnarray}

After a straightforward computation, we find for the variation of the action
\begin{eqnarray}
\delta I = -\,\int d^4x \biggl[\xi^k\,\Omega_k + (\partial_i\xi^k)\,\Omega_k{}^i + (\partial^2_{ij}\xi^k)\,\Omega_k{}^{ij} + (\partial^3_{ijn}\xi^k)\,\Omega_k{}^{ijn}\biggr],
\label{masterG} 
\end{eqnarray}
where explicitly
\begin{eqnarray}
\Omega_k &=& {\frac {\delta {\cal L}}{\delta g_{ij}}}\,\partial_kg_{ij} + {\frac {\delta {\cal L}}{\delta\psi^A}}\,\partial_k\psi^A + \,\partial_i\left({\frac {\partial {\cal L}}{\partial\partial_i\psi^A}} \,\partial_k\psi^A - \delta^i_k{\cal L} \right)\nonumber\\
&& + \,\partial_i\left({\frac {\partial {\cal L}}{\partial \partial_ig_{mn}}}\partial_kg_{mn}\right) + {\frac {\partial {\cal L}}{\partial \Gamma_{ln}{}^m}}\partial_k\Gamma_{ln}{}^m + {\frac {\partial {\cal L}}{\partial \partial_i\Gamma_{ln}{}^m}}\partial_k\partial_i\Gamma_{ln}{}^m,\label{Om1}\\
\Omega_k{}^i &=& 2{\frac {\delta {\cal L}}{\delta g_{ij}}}\,g_{kj} + {\frac {\delta {\cal L}}{\delta\psi^A}}\,(\sigma^A{}_B)_k{}^i\,\psi^B + {\frac {\partial {\cal L}}{\partial\partial_i\psi^A}}\partial_k\psi^A - \delta^i_k{\cal L} \nonumber\\ 
&& + \,2\partial_n\left({\frac {\partial {\cal L}}{\partial \partial_ng_{ij}}}g_{jk}\right) + {\frac {\partial {\cal L}}{\partial \partial_ig_{mn}}}\partial_kg_{mn} + \partial_j\!\left(\!{\frac {\partial {\cal L}}{\partial\partial_j\psi^A}}(\sigma^A{}_B)_k{}^i\psi^B\!\right)\!\nonumber\\
&& + \,{\frac {\partial {\cal L}}{\partial \Gamma_{li}{}^j}}\,\Gamma_{lk}{}^j + {\frac {\partial {\cal L}}{\partial \Gamma_{il}{}^j}}\,\Gamma_{kl}{}^j - {\frac {\partial {\cal L}}{\partial \Gamma_{lj}{}^k}}\,\Gamma_{lj}{}^i + {\frac {\partial {\cal L}}{\partial \partial_i\Gamma_{ln}{}^m}}\,\partial_k \Gamma_{nl}{}^m \nonumber\\
&& +\,{\frac {\partial {\cal L}}{\partial \partial_n\Gamma_{il}{}^m}} \,\partial_n\Gamma_{kl}{}^m  + \,{\frac {\partial {\cal L}}{\partial \partial_n\Gamma_{li}{}^m}}\,\partial_n \Gamma_{lk}{}^m - {\frac {\partial {\cal L}}{\partial \partial_n\Gamma_{lm}{}^k}}
\,\partial_n\Gamma_{lm}{}^i ,\label{Om2}\\
\Omega_k{}^{ij} &=& {\frac {\partial {\cal L}}{\partial\partial_{(i}\psi^A}} (\sigma^A{}_B)_k{}^{j)}\psi^B + {\frac {\partial {\cal L}}{\partial \Gamma_{(ij)}{}^k}}
+ {\frac {\partial {\cal L}}{\partial \partial_{(i}\Gamma_{j)l}{}^m}}\Gamma_{kl}{}^m \nonumber\\
&& + \,2{\frac {\partial {\cal L}}{\partial \partial_{(i}g_{j)n}}}g_{kn} + {\frac {\partial {\cal L}}{\partial \partial_{(i}\Gamma_{|l|j)}{}^m}}\,\Gamma_{lk}{}^m - {\frac {\partial {\cal L}}{\partial \partial_{(i}\Gamma_{|ln|}{}^k}}\,\Gamma_{ln}{}^{j)}. \label{Om3}\\ 
\Omega_k{}^{ijn} &=& {\frac {\partial {\cal L}}{\partial \partial_{(n}\Gamma_{ij)}{}^k}}.\label{Om4}
\end{eqnarray}
Invariance of the action, $\delta I = 0$, yields the four Noether identities:
\begin{equation}\label{NoeG}
\Omega_k = 0,\quad \Omega_k{}^i = 0,\quad \Omega_k{}^{ij} = 0, \quad \Omega_k{}^{ijn} = 0.
\end{equation}
General coordinate symmetry is due to the fact that the density ${\cal L}$ is constructed from covariant objects. Denoting 
$\rho^{ijk}{}_l = {\frac {\partial {\cal L}}{\partial R_{ijk}{}^l}},\,
\sigma^{ij}{}_k = {\frac {\partial {\cal L}}{\partial T_{ij}{}^k}},\,
\nu^{kij} = {\frac {\partial {\cal L}}{\partial Q_{kij}}},\,
\mu^{ij}{}_k = {\frac {\partial {\cal L}}{\partial N_{ij}{}^k}}$,
we find 
\begin{eqnarray}
{\frac {\partial {\cal L}}{\partial \Gamma_{ij}{}^k}} &=& -\,{\frac {\partial {\cal L}} {\partial \nabla_i\psi^A}}(\sigma^A{}_B)_k{}^j\,\psi^B + 2\nu^{ij}{}_k + 2\sigma^{ij}{}_k + 2\rho^{inl}{}_k\Gamma_{nl}{}^j + 2\rho^{nij}{}_l\Gamma_{nk}{}^l - \mu^{ij}{}_k,\label{dLG}\\
{\frac {\partial {\cal L}}{\partial \partial_i\Gamma_{jk}{}^l}} &=& 2\rho^{ijk}{}_l,\qquad {\frac {\partial {\cal L}}{\partial \partial_kg_{ij}}} = -\,\nu^{kij} + {\frac 12}\left(\mu^{(ki)j} + \mu^{(kj)i} - \mu^{(ij)k}\right).\label{dLdm}
\end{eqnarray}
As a result, we verify that $\Omega_k{}^{ij} = 0$ and $\Omega_k{}^{ijn} = 0$ are satisfied identically. Using (\ref{dLG}) and (\ref{dLdm}), we then recast the Noether identities (\ref{Om1}) and (\ref{Om2}) into
\begin{eqnarray}
\Omega_k &=& {\frac {\delta {\cal L}}{\delta g_{ij}}}\,\partial_kg_{ij} + {\frac {\delta {\cal L}}{\delta\psi^A}}\,\partial_k\psi^A + \partial_i\!\left(\!{\frac {\partial {\cal L}}{\partial\nabla_i\psi^A}}\nabla_k\psi^A - \delta^i_k{\cal L}\right)\nonumber\\
&& +\,\widehat{\nabla}{}_j\!\left(\!{\frac {\partial {\cal L}}{\partial\nabla_j\psi^A}} \,(\sigma^A{}_B)_m{}^n\,\psi^B\!\right)\!\Gamma_{kn}{}^m 
+ \,{\frac {\partial {\cal L}}{\partial\nabla_l\psi^A}}\,(\sigma^A{}_B)_m{}^n \,\psi^B\,R_{lkn}{}^m \nonumber\\
&& -\,\left[ \widehat{\nabla}{}_j\nu^{jmn} - {\frac 12}\check{\nabla}_i\left(\mu^{(im)n} + \mu^{(in)m} - \mu^{(mn)i}\right)\right]\partial_k g_{mn}\nonumber\\
&& + \,\rho^{iln}{}_m\partial_kR_{iln}{}^m + \sigma^{ln}{}_m\partial_kT_{ln}{}^m + \mu^{ln}{}_m\partial_kN_{ln}{}^m + \nu^{lmn}\partial_kQ_{lmn}=0,\label{Om1a}
\end{eqnarray}
\begin{eqnarray}
\Omega_k{}^i &=& 2{\frac {\delta {\cal L}}{\delta g_{ij}}}\,g_{kj} + {\frac {\delta {\cal L}}{\delta\psi^A}}\,(\sigma^A{}_B)_k{}^i\,\psi^B + {\frac {\partial {\cal L}}{\partial\nabla_i\psi^A}}\nabla_k\psi^A - \delta^i_k{\cal L} \nonumber\\ 
&& - \widehat{\nabla}{}_j\!\!\left(\!  2\nu^{ji}{}_k - {\frac {\partial {\cal L}}{\partial\nabla_j\psi^A}}(\sigma^A{}_B)_k{}^i\psi^B\!\right)\! +\check{\nabla}_n\left(\mu^{(ni)j} + \mu^{(nj)i} - \mu^{(ij)n}\right)g_{jk}\nonumber\\
&& -\,\mu^{ln}{}_kN_{ln}{}^i + \mu^{il}{}_nN_{kl}{}^n + \mu^{li}{}_nN_{lk}{}^n + 2\sigma^{il}{}_nT_{kl}{}^n - \sigma^{ln}{}_kT_{ln}{}^i\nonumber\\
&& + \,2\rho^{iln}{}_mR_{kln}{}^m + \rho^{lni}{}_mR_{lnk}{}^m - \rho^{lnm}{}_kR_{lnm}{}^i
 + \nu^{imn}Q_{kmn} = 0.\label{Om2a}
\end{eqnarray}
An arbitrary tensor density ${\cal A}^n{}_{i\dots}{}^{j\dots}$ is mapped into a density of the same weight by
\begin{equation}
\widehat{\nabla}{}_n{\cal A}^n{}_{i\dots}{}^{j\dots} = \partial_n{\cal A}^n{}_{i\dots}{}^{j\dots} + \Gamma_{nl}{}^j{\cal A}^n{}_{i\dots}{}^{l\dots} - \Gamma_{ni}{}^l
{\cal A}^n{}_{l\dots}{}^{j\dots},\label{dA}
\end{equation}
A similar covariant derivative, defined by the Riemannian connection, is denoted
\begin{equation}
\check{\nabla}{}_n{\cal A}^n{}_{i\dots}{}^{j\dots} = \partial_n{\cal A}^n{}_{i\dots}{}^{j\dots} + \widetilde{\Gamma}_{nl}{}^j{\cal A}^n{}_{i\dots}{}^{l\dots} - \widetilde{\Gamma}_{ni}{}^l
{\cal A}^n{}_{l\dots}{}^{j\dots},\label{dAc}
\end{equation}
It is worthwhile to note that the variational derivative with respect to the metric is an explicitly covariant density. This follows from the fact that the Lagrangian depends on $g_{ij}$ not only directly, but also through the objects $Q_{kij}$ and $N_{ki}{}^j$. Explicitly, we find
\begin{eqnarray}
{\frac {\delta {\cal L}}{\delta g_{ij}}} = {\frac {d {\cal L}}{d g_{ij}}} - \partial_n\left({\frac {\partial {\cal L}}{\partial \partial_ng_{ij}}}\right) 
= {\frac {\partial {\cal L}}{\partial g_{ij}}} + \widehat{\nabla}_n\nu^{nij} - {\frac 12}
\check{\nabla}_n\left(\mu^{(ni)j} + \mu^{(nj)i} - \mu^{(ij)n}\right).\label{dLgij}
\end{eqnarray}

The Noether identity (\ref{Om1a}) is apparently noncovariant in contrast to (\ref{Om2a}). To fix this, we replace $\Omega_k = 0$ by an equivalent covariant identity: $\overline{\Omega}{}_k = \Omega_k -  \Gamma_{kn}{}^m\Omega_m{}^n = 0$. Explicitly,
\begin{eqnarray}
\overline{\Omega}{}_k \!\!\!&=&\!\!\!{\frac {\delta {\cal L}}{\delta\psi^A}}\,\nabla_k\psi^A + \widehat{\nabla}{}_i\!\left(\!{\frac {\partial {\cal L}}{\partial\nabla_i\psi^A}} \,\nabla_k\psi^A - \delta^i_k{\cal L} \!\right)\! - \left({\frac {\partial {\cal L}}{\partial\nabla_i\psi^A}}\nabla_l\psi^A - \delta^i_l{\cal L}\right) T_{ki}{}^l\nonumber\\ 
&&\!\!\! +\left[ \widehat{\nabla}{}_n\nu^{nij} - {\frac 12}\check{\nabla}_n\left(\mu^{(ni)j} + \mu^{(nj)i} - \mu^{(ij)n}\right) - {\frac {\delta {\cal L}}{\delta g_{ij}}}\right]Q_{kij} + {\frac {\partial {\cal L}}{\partial\nabla_l\psi^A}}(\sigma^A{}_B)_m{}^n\psi^BR_{lkn}{}^m\nonumber\\
&&\!\!\! +\,\rho^{iln}{}_m\nabla_kR_{iln}{}^m +\,\sigma^{ln}{}_m\nabla_kT_{ln}{}^m 
+ \nu^{lmn}\nabla_k Q_{lmn} + \mu^{ln}{}_m\nabla_kN_{ln}{}^m = 0.\label{Om1b}
\end{eqnarray}

When the matter fields satisfy the field equations ${\delta {\cal L}}/{\delta\psi^A} = 0$, the Noether identities (\ref{Om2a}) and (\ref{Om1b})  reduce to the {\it conservation laws}  for the energy-momentum and hypermomentum, respectively. 

{\bf Nonminimal coupling models} \cite{Obukhov:Koivisto:2006,Obukhov:Bertolami:2007,Obukhov:Mohseni:2010} have attracted considerable attention recently. Using our results above, we can analyse a large class of models with the Lagrangian
\begin{equation}
{\cal L} = \sqrt{-g}FL_{\rm mat}.\label{Lnon}
\end{equation}
The coupling function $F = F(g_{ij},R_{kli}{}^j,T_{kl}{}^i, Q_{kij}, N_{kl}{}^i)$ depends arbitrarily on its arguments, whereas the matter Lagrangian $L_{\rm mat} = L_{\rm mat}(\psi^A, \nabla_i\psi^A, g_{ij})$ has the usual form. 

The matter is characterized by the canonical energy-momentum tensor, the canonical hypermomentum tensor, and the metrical energy-momentum tensor
\begin{equation}
\Sigma_k{}^i = {\frac {\partial {L_{\rm mat}}}{\partial\nabla_i\psi^A}} \,\nabla_k\psi^A - \delta^i_kL_{\rm mat},\quad
\Delta^n{}_k{}^i = -\,{\frac {\partial {L_{\rm mat}}}{\partial\nabla_i\psi^A}} \,(\sigma^A{}_B)_k{}^n \psi^B,\quad
t_{ij} = {\frac 2{\sqrt{-g}}}\,{\frac {\delta {(\sqrt{-g}L_{\rm mat})}}{\delta g^{ij}}}.
\end{equation}
The usual spin arises as an antisymmetric part of the hypermomentum, $\tau_{ij}{}^k = \Delta_{[ij]}{}^k$, whereas the trace $\Delta^k = \Delta^i{}_i{}^k$ is the dilation current. The symmetric traceless part describes the proper hypermomentum \cite{Obukhov:Hehl:1995}.

The conservation laws are derived from (\ref{Om2a}) and (\ref{Om1b}) , and they read 
\begin{eqnarray}\label{cons1b}
F\Sigma_k{}^i &=& Ft_k{}^i + {\stackrel * \nabla}{}_n\left(F\Delta^i{}_k{}^n\right),\\
{\stackrel * \nabla}{}_i\left(F\Sigma_k{}^i\right) &=& F \left( \Sigma_l{}^i T_{ki}{}^l 
- \Delta^m{}_n{}^l R_{klm}{}^n  - {\frac 12}t^{ij}Q_{kij} \right)- L_{\rm mat}\nabla_kF.\label{cons2b}
\end{eqnarray}
The so-called modified covariant derivative is defined as ${\stackrel * \nabla}{}_i = \nabla_i + N_{ki}{}^k$. These results generalize our previous findings \cite{Obukhov:Puetzfeld:2008,Obukhov:Puetzfeld:2013:1,Obukhov:Puetzfeld:2013:2}. 

{\bf The equations of motion} of extended bodies are obtained from the conservation laws, see the historic overview in \cite{Obukhov:Puetzfeld:2007}. There exist various schemes using the so-called multipole expansion technique in which the motion of an extended body, sweeping a finite world tube, is approximated by the motion of a point particle, which is characterized by a (infinite, in general) set of moments. The latter are defined as integrated quantities derived from the Noether currents that describe body's matter. Here we use the covariant expansion approach of Synge \cite{Obukhov:Synge:1960,Obukhov:Dixon:1964}.

In Synge's formalism, two-points tensors (or bitensors) are introduced as tensorial functions of two spacetime points. Most important among them is the world-function $\sigma(x,y)$, which measures the interval (distance) along a unique geodesic curve connecting the two points $x$ and $y$, and the parallel propagator $g^y{}_x(x,y)$ that transfers tensorial objects along this geodesic. Covariant derivatives of the world-function are denoted by $\sigma_y := \nabla_y\sigma$, etc.

Let us consider, for now, the special case when the microstructure of matter is reduced to the spin $\tau_{ij}{}^k$ and the geometry of spacetime, accordingly, is characterized by the vanishing nonmetricity $Q_{kij} = 0$. The general equations of motion based on the conservation laws (\ref{cons1b}) and (\ref{cons2b}) will be analysed elsewhere. 

The lowest (pole and dipole) integrated moments are $p^{y_0} = \int_{\Sigma(s)}g^{y_0}{}_{x_0}\widetilde{\Sigma}^{x_0 x_2} d \Sigma_{x_2}$, and
\begin{equation}
p^{y_1y_0} = - \int_{\Sigma(s)}\sigma^{y_1}g^{y_0}{}_{x_0}\widetilde{\Sigma}^{x_0 x_2} d \Sigma_{x_2},\qquad
s^{y_0y_1} = - \int_{\Sigma(s)}g^{y_0}{}_{x_0}g^{y_1}{}_{x_1}\widetilde{\tau}^{[x_0 x_1]x_2} d \Sigma_{x_2}.
\end{equation}
The tilde denotes densitized canonical energy-momentum and spin tensors, and the integration is done over the spatial cross-section $\Sigma(s)$ of a world tube of a body at the value $s$ of the proper time parameter along the representative world line $x^i(s)$.

Performing the appropriate integrations of the conservation laws (\ref{cons1b}) and (\ref{cons2b}), we obtain the equations of motion in the pole-dipole approximation \cite{Obukhov:Puetzfeld:2013:3}
\begin{equation}
{\frac{D}{ds}}{\cal P}^{a} = {\frac 12}\widetilde{R}^a{}_{bcd}{\cal J}^{cd}v^b + f^a,\qquad
{\frac{D}{ds}}{\cal J}^{ab} = - 2v^{[a}{\cal P}^{b]} + f^{ab}.\label{MP}
\end{equation}
Here $v^a = dx^a/ds$ is the 4-velocity of the body, and we construct the generalized total energy-momentum vector and the total angular momentum tensor 
\begin{eqnarray}
{\cal P}^a = F\left(p^a - {\frac 12}N^a{}_{cd}S^{cd}\right) + \left(p^{ba} - S^{ab}\right)\nabla_bF,\qquad {\cal J}^{ab} = F\left(L^{ab} + S^{ab}\right),\label{Jab}
\end{eqnarray}
{}from the integrated 4-momentum $p^a$ of the body, the integrated orbital angular momentum $L^{ab} = 2p^{[ab]}$, and the integrated spin angular momentum $S^{ab} = -2s^{ab}$.

The Mathisson-Papapetrou equations (\ref{MP}) contain an additional force and torque due to the higher multipole moments and the nonminimal coupling (with $A_i = \nabla_i\log F$):
\begin{eqnarray}
f^a &=& F\Theta^{bc}{}_d\widetilde{\nabla}{}^a T_{bc}{}^d - 2q^{bcd}N_{dc}{}^a\nabla_bF + 2Fq^{acd}\nabla_dA_c -\xi\nabla^aF + \xi^{b}\widetilde{\nabla}_b\nabla^aF,\label{fa}\\
f^{ab} &=& 2F\Theta^{cd[a}T_{cd}{}^{b]} + 4F\Theta^{[a}{}_{cd}T^{b]cd} - 4q^{[a|c|b]}\nabla_cF - 2\xi^{[a}\nabla^{b]}F.\label{fab}
\end{eqnarray}
Here $\xi = \int_{\Sigma(s)}\widetilde{L}_{\rm mat}w^{x_2}d\Sigma_{x_2}$, $\xi^y = \int_{\Sigma(s)}\sigma^y\widetilde{L}_{\rm mat}w^{x_2}d\Sigma_{x_2}$, $\Theta^{bca} = {\frac 12}\left(q^{bca} + q^{bac} - q^{cab}\right)$, and 
\begin{equation}
q^{y_0y_1y_2} = \int_{\Sigma(s)}g^{y_0}{}_{x_0}g^{y_1}{}_{x_1}g^{y_2}{}_{x_2}\widetilde{\tau}^{[x_0 x_1]x_2} w^{x_3}d\Sigma_{x_3}.\label{q}
\end{equation}
For the definition of $w^x$ see \cite{Obukhov:Dixon:1964}.

Interestingly, the form of the torsion-dependent pieces of the additional force and torque exactly reproduces the contribution of the quadrupole translational moment studied for fermionic matter in \cite{Obukhov:Hehl:1998,Obukhov:1998}. An important next step would be to establish the complete structure of the equations of motion up to the quadrupole order both in the rotational and translational moments. Such a study can be most conveniently done along the lines of the approach of Bailey and Israel \cite{Obukhov:Bailey:1975}.

Our covariant equations of motion (\ref{MP}) extend and confirm previous results on the dynamics of extended bodies with spin \cite{Obukhov:Stoeger:1979,Obukhov:Stoeger:1980} and \cite{Obukhov:Puetzfeld:2007}. In particular, when the coupling is minimal $(F = 1)$, we immediately verify that the post-Riemannian geometrical structure of spacetime can be detected only by using test particles with {\it intrinsic spin}. Rotating macroscopic bodies are thus, so to say, neutral to the torsion. 

Nevertheless, it is worthwhile to notice that even structureless massive point particles can be affected by the post-Riemannian gravitational field when the coupling function $F$ depends on the torsion and nonmetricity. Such single-pole particles do not move along geodesic curves (in contrast to minimally coupled point particles). A ``pressure'' like force arises as the gradient of the coupling function:
\begin{equation}
m\dot{v}^a = \xi\left(\delta^a_b - v^av_b\right)\nabla^b\log F.\label{logF}
\end{equation}
A similar force determines the nongeodetic motion of test particles in the scalar-tensor theory of gravitation \cite{Obukhov:Brans:1961,Obukhov:Fujii:2003} where the gravitational coupling constant is replaced by the scalar coupling function.

{\bf Acknowledgements} D.P. was supported by the Deutsche Forschungsgemeinschaft (DFG) through the grant LA-905/8-1/2.

\end{document}